\hsize = 5.20in
\hoffset = 0.15in
\vsize = 8.50in
\voffset = 0.25in
\magnification=1200



\def\fullpar#1#2{\leftskip#1\rightskip#2}

\def\outdent#1#2{\llap{\hbox to #1{#2 \hss}}\ignorespaces}
\def\parbegin#1#2#3#4{
	\fullpar{#1}{#2}
	\ifcase #3 \baselineskip = 1.5\baselineskip
	\or \baselineskip = 2\baselineskip
	\or \baselineskip = 3\baselineskip
	\else \baselineskip = 1\baselineskip\fi
	\ifdim #4 > 0in \else \noindent \fi
	\ignorespaces}
\overfullrule=0pt
\centerline{  }
\baselineskip=1.5 \baselineskip
\line{\hbox{ }\hfill \hbox{29th July 1996 }}
\vskip 2cm  
\centerline{\bf A CO-VARIANT APPROACH TO ASHTEKAR'S}
\centerline{\bf CANONICAL GRAVITY}
\vskip 1.2cm
\centerline{Brian P. Dolan {\it and} Kevin P. Haugh}
\vskip .5cm
\centerline{\it Department of Mathematical Physics, St. Patrick's College}
\centerline{\it Maynooth, Ireland}
\vskip .5cm
\centerline{e-mail: bdolan@thphys.may.ie, khaugh@thphys.may.ie}
\vskip 1.5cm
\centerline{ABSTRACT}
\noindent A Lorentz and general co-ordinate co-variant form
of canonical gravity, using Ashtekar's variables, is 
investigated. A co-variant treatment due to Crnkovic and
Witten is used, in which a point in phase space represents
a solution of the equations of motion and a symplectic
functional two form is constructed which is Lorentz
and general co-ordinate invariant. The subtleties and 
difficulties due to the
complex nature of Ashtekar's variables are addressed and
resolved.
\vfill\eject

\noindent {\bf 1. Introduction}
\vskip 5mm
In 1986, Abhay Ashtekar [1] discovered a set of canonical 
variables for the gravitational field as described by the 
general theory of relativity. Ashtekar found that they led 
to a considerable simplification of the constraints associated 
with the Hamiltonian formulation of Einstein's theory. Indeed, 
Ashtekar's constraints are polynomials in the canonical 
variables. Ashtekar's canonical gravity is definite progress in 
the 
direction of a quantum theory of gravity since it gives rise 
to a closed constraint algebra [2].

Hamiltonian models of physical phenomena have always 
distinguished between time and space. The Hamiltonian of 
a dynamical system generates time translations, that is 
to say it determines the time evolution of the dynamical 
variables. Relativity regards time and space as being 
components of a single entity : space-time. An equation, 
describing the way a physical quantity changes with time, 
does not look the same to all relativistic observers. In 
other words, an equation of this kind is not co-variant.
It is usual to develop the Hamilton mechanics of a 
relativistic field by specifying a space-time foliated 
by space-like hyper-surfaces of constant time, and a 
Hamiltonian functional on this space-time. However, this 
approach spoils co-variance from the beginning because a 
time co-ordinate must be singled out, in order for the 
required foliation to make sense [3]. 

One way of viewing the role of canonical variables is that their 
initial values determine a solution of the Hamilton equations. 
In other words, there is a one-to-one correspondence between the 
canonical variables at any time, $t$, and the canonical variables 
initially [4]. Thus we can describe the phase space as the set of 
solutions of the Hamilton equations of motion. 
For a field theory, a knowledge of the initial canonical variables
requires a knowledge of the field configuration and its time 
derivatives on a space-like hyper-surface, and a point in phase 
space is a solution of the Hamilton equations at a given time. 
The object of this paper is to describe Ashtekar's gravity in a 
manifestly co-variant way. One possible way of achieving this goal
is to use a simple construction due to Crnkovic and Witten. 

The essence of the Crnkovic-Witten construction is the observation
that a co-variant theory must have an invariant symplectic form, 
and that each point in phase space represents a solution of the 
equations of motion. One can thus dispense with the Hamiltonian, 
and focus on the symplectic structure and the points of phase 
space as providing a co-variant description of the dynamics in 
phase space. 
This idea has been successfully applied by Crnkovic and 
Witten [5] to the Yang-Mills field and to general relativity 
(using the 3-metric and the extrinsic curvature as canonical 
variables) where there is an additional complication due to 
gauge invariance.

It is not immediately obvious how to implement the Crnkovic-Witten
construction in the framework of Ashtekar's canonical gravity. In 
particular, the complex nature of the canonical variables leads 
to difficulties which will be addressed here. It will be shown 
that these difficulties can be overcome, and the Crnkovic-Witten 
construction can be applied successfully to give a co-variant 
version of Ashtekar's theory.

\vskip 5mm
\noindent {\bf 2. Ashtekar's Canonical Gravity}
\vskip 5mm

In this section, we shall review Ashtekar's Hamiltonian 
formulation with a view to establishing our notation and 
conventions.
Ashtekar's canonical variables are the inverse 
densitized triads, 
$E^{ai}$, and the Ashtekar connection, $A_{ai}$, defined on a 
space-like hyper-surface, $\Sigma_{t}$, of constant time, $t$. 
(Ashtekar's canonical variables can also be defined on a null 
hyper-surface [6].) Here $a$ and $i$ are orthonormal and 
co-ordinate indices respectively, ranging from $1$ to $3$. 
The metric signature is $-+++$, and the completely anti-symmetric 
Levi-Civita tensor is taken to be  $\varepsilon_{0123} = 1$. 
For a space-like foliation, a set of orthonormal $1$-forms is 
given by
$$\openup 2mm
e^{0}\quad =\quad N\enspace dt,\qquad
e^{a}\quad =\quad 
h^{a}{}_{i}N^{i}\enspace dt 
+ h^{a}{}_{i}\enspace d{x^i},\eqno (1)
$$
\noindent where $N$ and ${N^i}$ are the lapse and shift functions 
respectively. The dual basis vectors are
$$\openup 2mm
\beta{_0}\quad =\quad
{1\over N}\left({\partial\over\partial{t}}
- {N^i}{\partial\over\partial{x^i}}\right), \qquad
\beta{_a}\quad =\quad 
(h^{-1})_{a}{}^{i}{\partial\over\partial{x^i}}.\eqno (2) 
$$
\noindent Each $\beta_a$ is space-like, and the normal 
$\beta_0$ is time-like. The {\it densitized triads} are 
defined by
$$\openup 2mm
(E^{-1})_{ai}\quad =\quad{1\over h} h_{ai},\eqno (3)
$$
\noindent where $h$ is the determinant of the matrix, 
$[h_{ai}]$. The densitized triads are real-valued on 
any co-ordinate patch provided that $h^{0}{}_{i} = 0$ [7]. 
This is the {\it time gauge} condition, which can be relaxed 
by allowing the densitized triads to become complex-valued 
(see the appendix). The local group of local tangent space 
rotations, that preserves the time gauge condition, is the 
rotation group, $SO(3)$. The {\it inverse densitized triads}, 
$E^{ai}$, satisfy 
$$\openup 2mm
E^{ai}(E^{-1})_{aj}\quad =\quad\delta^{i}{}_{j}.\eqno (4)
$$
\noindent Let $E$ be the determinant of the matrix, $[E^{ai}]$. 
Now 
$$\openup 2mm
E^{ai}\quad =\quad h (h^{-1})^{ai},\qquad 
E\quad =\quad{h^2}.\eqno (5)
$$
\noindent We record the useful relations :
$$\openup 2mm
h_{ai}\quad =\quad{\sqrt E}(E^{-1})_{ai},\qquad
(h^{-1})^{ai}\quad =\quad{E^{ai}\over{\sqrt E}}.\eqno (6)
$$
\noindent The torsion-free, metric-compatible connection $1$-forms
$\omega_{_{AB}}$ are given by
$$\openup 2mm
\omega_{_{AB}}\quad =
\quad{1\over 2}\left [\bigl ({i_{_A}}{i_{_B}}d{e_{_C}}\bigr )
{e^{^C}}
- {i_{_A}}d{e_{_B}} + {i_{_B}}d{e_{_A}}\right ],\eqno (7)
$$
where $A,B,\ldots$ range from $0$ to $3$, ${i_{_A}}$ is the 
interior derivative along $\beta_{_{A}}$, and $d$ is the exterior 
derivative. The {\it Ashtekar connection}, $A_{ai}$, can then be 
defined in terms of the components of the connection $1$-forms, 
$\omega_{_{AB}}$ : 
$$\openup 2mm
A_{ai}\quad =\quad\omega_{0ai} -{i\over 2}
\varepsilon_{abc}\omega^{bc}{}_{i}.\eqno (8)
$$
\noindent The curvature $2$-forms, $R_{_{AB}}$, are given by
$$\openup 2mm\eqalignno{  
R_{_{AB}}\quad &=\quad d\omega_{_{AB}}
+\omega_{_{AC}}\wedge\omega^{^C}{}_{_B},\quad
R_{_{BA}}\quad =\quad - R_{_{AB}}. & (9)\cr
\noalign{\hbox{They satisfy the Hodge duality relations :}}
*R_{_{AB}}\quad &=\quad{1\over 2}\varepsilon_{_{ABCD}}
R^{^{CD}},\quad **R_{_{AB}}\quad =\quad - R_{_{AB}}.& (10)\cr
\noalign{\hbox{The self-dual curvature $2$-forms, $^{+}R_{_{AB}}$,
are then given by}}
^{+}R_{_{AB}}\quad &=\quad R_{_{AB}} -i *R_{_{AB}},\quad
*^{+}R_{_{AB}}\quad =\quad i ^{+}R_{_{AB}}. & (11)\cr
}$$

\noindent In the absence of torsion, we have the identity :
$$\openup 2mm
R_{_{AB}}\wedge{e^{^B}}\quad =\quad 0.\eqno (12)
$$
\noindent Thus, for a vacuum gravitational field, the action 
density $4$-form can be written :
$$\openup 2mm\eqalignno{  
{\cal L}{d^4}x\quad &=\quad{1\over 2}R_{_{AB}}\wedge *e^{^{AB}}\cr
\quad &=\quad{1\over 2} *R_{_{AB}}\wedge e^{^{AB}}
+ {i\over 2}R_{_{AB}}\wedge e^{^{AB}}\cr
\quad &=\quad{i\over 2}\enspace ^{+}R_{_{AB}}\wedge e^{^{AB}}, 
& (13)\cr 
\noalign{\hbox{where $e^{^{AB}}$ stands for $e^{^A}\wedge e^{^B}$.
Now (11) tells us}}
^{+}R^{bc}\quad &=\quad i\varepsilon^{abc}\enspace ^{+}R_{0a}, 
& (14)\cr
\noalign{\hbox{and therefore,}}
{\cal L}{d^4}x\quad &=\quad
i ^{+}R_{0a}\wedge\left (e^{0a} +{i\over 2}\varepsilon^{abc}
e_{bc}\right). & (15)\cr
}$$

\noindent Writing
$$\openup 2mm
F_{a}\quad =\quad ^{+}R_{0a},\quad 
\Lambda^{a}\quad =\quad
e^{0a} +{i\over 2}\varepsilon^{abc}e_{bc},\eqno (16)
$$
\noindent we have
$$\openup 2mm
{\cal L}{d^4}x\quad =\quad iF_{a}\wedge\Lambda^{a}.\eqno (17)
$$
\noindent It is straightforward to obtain the important relations :
$$\openup 2mm\eqalignno{
F_{a}\quad =&\quad{1\over 2} F_{a\mu\nu} d{x^\mu}\wedge d{x^\nu}\cr
\quad =&\quad\left({\dot A}_{ai} -\partial_{i} A_{a0}
 - i\varepsilon_{abc}A^{b}{}_{0}A^{c}{}_{i}\right) dt\wedge d{x^i}
\cr & + {1\over 2}\left(\partial_{i} A_{aj} - \partial_{j} A_{ai} 
- i\varepsilon_{abc}A^{b}{}_{i}A^{c}{}_{j}\right) 
d{x^i}\wedge d{x^j}, & (18)\cr
\Lambda^{a}\quad =&\quad{1\over 2}\Lambda^{a}{}_{\mu\nu} d{x^\mu}
\wedge d{x^\nu}\cr 
\quad =&\quad {1\over 2}\left (\varepsilon^{abc}{\cal N}
E_{b}{}^{j}E_{c}{}^{k} + 2iE^{aj}{N^k}\right)
\varepsilon_{ijk}dt\wedge d{x^i}\cr
& +{i\over 2}\varepsilon_{ijk}E^{ai} d{x^j}\wedge d{x^k},
& (19)\cr
}$$
\noindent with the help of (1), (6), (8), and (11). Here, we have 
used the symbol ${\cal N}$ for $N/{\sqrt E}$, and the Greek letters
$\mu,\nu,\ldots$ for co-ordinate indices, ranging from $0$ to $3$.
\noindent It follows that
$$\openup 2mm\eqalignno{
{\cal L}{d^4}x\quad =\quad
\biggl [&A_{ai}\dot{E}^{ai} 
- A_{a0}\bigl (\partial_{i} E^{ai}
- i\varepsilon^{abc}A_{bi}E_{c}{}^{i}\bigr )\cr
& +{i\over 2}{\cal N}\varepsilon^{abc}
F_{aij}E_{b}{}^{i}E_{c}{}^{j} 
+ N^{i}F_{aij}E^{aj}\biggr ] dt\wedge{d^3}x. & (20)\cr
}$$
 
\noindent Thus we see that the Ashtekar connection, $A_{ai}$, are 
the momenta conjugate to the inverse densitized triads, $E^{ai}$, 
and that Ashtekar's Hamiltonian is
$$\openup 2mm
H\quad =\quad\int_{\Sigma_t}{d^3}x
\biggl [A_{a0}\bigl (\partial_{i} E^{ai}
- i\varepsilon^{abc}A_{bi}E_{c}{}^{i}\bigr)
-{i\over 2}{\cal N}\varepsilon^{abc}
F_{aij}E_{b}{}^{i}E_{c}{}^{j} 
- N^{i}F_{aij}E^{aj}\biggr ],\eqno (21)
$$
\noindent for a space-like hyper-surface, $\Sigma_t$. This is 
the form of the
Hamiltonian given in [8].
 
The general theory of relativity has a phase space structure 
analogous to that of the $SU(2)$ Yang-Mills field, where local 
$SO(3)$ 
tangent space rotations, or more generally, Lorentz transformations
play the role of gauge transformations in Yang-Mills theory. 
General co-ordinate 
invariance and Lorentz invariance require the introduction of 
redundant canonical variables. This leads to constraints 
expressing the resulting interdependence of the canonical 
variables. 
 
In Ashtekar's formulation, the constraints take the polynomial 
form :
$$\openup 2mm\eqalignno{  
{\delta H\over\delta A_{a0}}\quad &=\quad
\partial_{i} E^{ai} - i\varepsilon^{abc}
A_{bi}E_{c}{}^{i}\quad =\quad 0, &\cr
{\delta H\over\delta{\cal N}}\quad &=\quad
- {i\over 2}\varepsilon^{abc}F_{aij}E_{b}{}^{i}E_{c}{}^{j} 
\quad =\quad 0, &\cr
{\delta H\over\delta{N^i}}\quad &=\quad
-F_{aij}E^{aj}\quad =\quad 0, & (22)\cr
}$$
everywhere on the hyper-surface, $\Sigma_t$.
 
Ashtekar [9] has shown that these secondary constraints are 
first-class. We see that the Hamiltonian is a linear combination 
of the constraints. It is therefore first-class and weakly zero. 
It is important to recall that the Yang-Mills Hamiltonian is not 
weakly zero in general. This reflects a dynamical difference 
between the Yang-Mills field and the gravitational field.

\vskip 5mm
\noindent {\bf 3. Crnkovic-Witten Theory}
\vskip 5mm
Let us review the Crnkovic-Witten construction in the case of 
the scalar field. We begin with the action of the scalar field in 
flat space-time :
$$\openup 2mm\eqalignno{
S\quad &=\quad\int_{M}{d^4}x\enspace{\cal L},\cr
{\cal L}\quad &=\quad{1\over 2}\left(
\partial_{\mu}\phi\partial^{\mu}\phi - V(\phi)\right). & (23)\cr
}$$
 
\noindent Crnkovic and Witten's idea involves the introduction 
of a {\it symplectic current} at each spacetime point, $x$ :
$$\openup 2mm
J_{\mu}(x)\quad =\quad\delta\left(
{\delta{\cal L}\over{\delta({\partial_\mu}\phi)}}\right)
\wedge\delta\phi(x)\quad =\quad
\delta\partial_{\mu}\phi(x)\wedge\delta\phi(x),\eqno (24)
$$
\noindent where $\delta$ stands for the functional exterior 
derivative of forms on the phase space of the scalar field [5]. Now
$$\openup 2mm\eqalignno{
\delta J_{\mu}(x)\quad &=\quad\delta\left(\delta
\partial_{\mu}\phi(x)\right)\wedge\delta\phi(x)
-\delta\partial_{\mu}\phi(x)\wedge
\delta\left(\delta\phi(x)\right)\cr
\quad &=\quad\partial_{\mu}\delta\left(\delta\phi(x)\right) 
\quad =\quad 0. & (25)\cr
}$$
\noindent This means that $J_{\mu}$ is closed as a functional 
$2$ -form. Further,
$$\openup 2mm\eqalignno{
\partial^{\mu}J_{\mu}(x)\quad &=\quad
\delta\left(\partial^{\mu}\partial_{\mu}\phi(x)\right)\wedge
\delta\phi(x)
+\delta\partial_{\mu}\phi(x)\wedge\delta\partial^{\mu}\phi(x)\cr
\quad &=\quad - V''(\phi)\delta\phi(x)\wedge\delta\phi(x)
+\delta\partial_{\mu}\phi(x)\wedge\delta\partial^{\mu}\phi(x)\cr
\quad &=\quad 
-\delta\partial_{\mu}\phi(x)\wedge\delta\partial^{\mu}\phi(x)
\quad =\quad 0, & (26)\cr
}$$
\noindent with the help of the equation of motion
$$\openup 2mm
\partial_{\mu}\partial^{\mu}\phi + V'(\phi)\quad =\quad 0.
\eqno (27)
$$
\noindent Stokes' theorem then implies that
$$\openup 2mm
\int_{N}{d^4}x\enspace\partial^{\mu}J_{\mu}(x)\quad =\quad
\int_{\partial N}d\sigma^{\mu}(x)J_{\mu}(x)\quad =\quad 0,
\eqno (28)
$$
\noindent where $N$ is a sub-manifold of $M$ with boundary, 
$\partial N$. Suppose ${\partial N} =\Sigma_{t_1}\cup\Sigma_{t_2}
\cup\Sigma$, where $\Sigma_{t_1},\Sigma_{t_2}$ are space-like 
hyper-surfaces of constant time, and $d\sigma^{\mu}J_{\mu}$ 
vanishes everywhere on the hyper-surface, $\Sigma$. Then
$$\openup 2mm
\int_{\Sigma_{t_1}}d\sigma^{\mu}(x)J_{\mu}(x)\quad =
\quad\int_{\Sigma_{t_2}}d\sigma^{\mu}(x)J_{\mu}(x),\eqno (29)
$$
\noindent where $d\sigma^{\mu}$ is chosen to point in the 
same temporal direction on both $\Sigma_{t_1}$ and 
$\Sigma_{t_2}$. This means that the closed functional $2$-form 
$$\openup 2mm\eqalignno{
\Omega\quad &=\quad\int_{\Sigma_t}d\sigma^{\mu}(x)J_{\mu}(x)\cr
\quad &=\quad\int_{\Sigma_t}d\sigma^{\mu}(x)
\delta\partial_{\mu}\phi(x)\wedge\delta\phi(x) & (30)\cr
}$$
\noindent is independent of the choice of $\Sigma_{t}$. When we 
perform a Lorentz transformation $\Sigma_{t}\to\Sigma_{t'}$ and 
$\Omega\to\Omega'$, where
$$\openup 2mm
\Omega'\quad =\quad
\int_{\Sigma_{t'}}d\sigma^{\mu}(x')J_{\mu}(x')
\quad =\quad\int_{\Sigma_t}d\sigma^{\mu}(x)J_{\mu}(x)
\quad =\quad\Omega.\eqno (31)
$$
 
We conclude that $\Omega$ is a Lorentz invariant symplectic 
form on the phase space of the scalar field, and that it is 
possible to formulate the Hamiltonian theory of the scalar 
field in a manifestly co-variant way. The Lorentz invariance 
of the symplectic form allows us to choose a space-like 
hyper-surface, $\Sigma_{t}$, such that 
$$\openup 2mm
d\sigma^{0}(x)\quad =\quad{d^3}x,
\qquad d\sigma^{i}(x)\quad =\quad 0\eqno (32)
$$
\noindent for all $x\in\Sigma_{t}$, and hence, we obtain the 
standard symplectic form
$$\openup 2mm
\Omega\quad =\quad\int_{\Sigma_{t}}{d^3}x\enspace
\delta{\dot\phi}(x)\wedge\delta\phi(x)\quad =\quad
\int_{\Sigma_{t}}{d^3}x\enspace
\delta\left({\delta{\cal L}\over
\delta{\dot\phi}(x)}\right)\wedge\delta\phi(x),\eqno (33)
$$
\noindent where ${\dot\phi}$ is the momentum canonically 
conjugate to $\phi$. 
 
Next we consider the construction of a Lorentz invariant and 
gauge invariant symplectic form on the phase space of the 
$SU(2)$ Yang-Mills field, $A_{\mu}$, in flat space-time. In 
this case, the action is 
$$\openup 2mm\eqalignno{
S\quad &=\quad - {1\over 4}\int_{M}{d^4}x\enspace tr 
(F_{\mu\nu}F^{\mu\nu}), &\cr
\noalign{\hbox{where}}
F_{\mu\nu}\quad &=\quad
\partial_{\mu}A_{\nu} -\partial_{\nu}A_{\mu} 
+ [A_{\mu}, A_{\nu}]. & (34)\cr
}$$
\noindent The symplectic current is taken to be
$$\openup 2mm\eqalignno{
J_{\mu}\quad &=\quad tr\enspace\delta\left({\delta{\cal L}
\over{\delta({\partial_\mu}{A_\nu})}}\right)\wedge
\delta A^{\nu}\quad =\quad 
tr\enspace\delta F_{\mu\nu}\wedge\delta A^{\nu}, & (35)\cr
}$$
where $\delta$ is the functional exterior derivative of forms on 
the Yang-Mills phase space [5].
This symplectic current is closed, since
$$\openup 2mm\eqalignno{
\delta\left(\delta F_{\mu\nu}\right)\quad &=\quad 0,\qquad
\delta\left(\delta A^{\mu}\right)\quad =\quad 0, & (36)\cr
\noalign{\hbox{and therefore,}}
\delta J_{\mu}\quad &=\quad tr\enspace\delta
\left(\delta F_{\mu\nu}\right)\wedge\delta A^{\nu} - 
tr\enspace\delta F_{\mu\nu}\wedge\delta\left(\delta A^{\nu}\right)
\quad =\quad 0. & (37)\cr
}$$
 
On introducing a basis, say $\{T^{a}\}$, for the $SU(2)$ 
Lie algebra, we have 
$$\openup 2mm\eqalignno{
\nabla_{\mu} A_{a\nu}\quad &=\quad\partial_{\mu}A_{a\nu} +
[A_{\mu}, A_{\nu}]_{a}\cr
\quad &=\quad\partial_{\mu}A_{a\nu} -\varepsilon_{abc}
A^{b}{}_{\mu}A^{c}{}_{\nu}. & (38)\cr
}$$ 
\noindent Thus
$$\openup 2mm\eqalignno{
\delta F_{a\mu\nu}\quad &=\quad\nabla_{\mu}\delta A_{a\nu} 
-\nabla_{\nu}\delta A_{a\mu}, &\cr
\noalign{\hbox{or}}
\delta F_{\mu\nu}\quad &=\quad\nabla_{\mu}\delta A_{\nu} -
\nabla_{\nu}\delta A_{\mu}. & (39)\cr
}$$
\noindent As a result of a gauge transformation, $A_{a\mu}\to 
A'_{a\mu}$ with
$$\openup 2mm\eqalignno{
A'_{a\mu}\quad &=\quad A_{a\mu}
+\partial_{\mu}\lambda_{a} + [A_{\mu},\lambda]_{a} & (40)\cr
\noalign{\hbox{for some infinitesimal real-valued function 
$\lambda$ on space-time. We have}}
\delta A'_{a\mu}\quad &=\quad 
\delta A_{a\mu} + [\delta A_{\mu},\lambda]_{a}, & (41)\cr
\delta F'_{a\mu\nu}\quad &=\quad 
\delta F_{a\mu\nu} + [\delta F_{\mu\nu},\lambda]_{a}. & (42)\cr
}$$
\noindent The symplectic current transforms according to
$$\openup 2mm\eqalignno{
J'_{\mu}\quad &=\quad\delta F'_{a\mu\nu}\wedge\delta A'^{a\nu}\cr
\quad &=\quad J_{\mu} -\varepsilon_{abc}\lambda^{c}\left(
\delta F^{b}{}_{\mu\nu}\wedge\delta A^{a\nu}
+\delta F^{a}{}_{\mu\nu}\wedge\delta A^{b\nu}\right) + O
\left({\lambda^2}\right)\cr
\quad &=\quad J_{\mu} + O\left({\lambda^2}\right). & (43)\cr
}$$
\noindent Thus the symplectic current is an $SU(2)$-singlet. 
This allows us to write
$$\openup 2mm\eqalignno{
\partial^{\mu}J_{\mu}\quad &=\quad\nabla^{\mu}J_{\mu}\cr
\quad &=\quad 
tr\enspace\nabla^{\mu}\delta F_{\mu\nu}\wedge\delta A^{\nu}
+ tr\enspace\delta F_{\mu\nu}\wedge\nabla^{\mu}\delta A^{\nu}. 
& (44)\cr
}$$
\noindent The equations of motion   
$$\openup 2mm\eqalignno{
\nabla^{\mu}F_{\mu\nu}\quad &=\quad\partial^{\mu}F_{\mu\nu}
+ [A^{\mu}, F_{\mu\nu}]\quad =\quad 0, & (45)\cr
\noalign{\hbox{imply that}} 
\nabla^{\mu}\delta F_{\mu\nu}\quad &=\quad 
- [\delta A^{\mu}, F_{\mu\nu}]. & (46)\cr
}$$
\noindent Also
$$\openup 2mm\eqalignno{
tr\enspace\nabla^{\mu}\delta F_{\mu\nu}\wedge\delta A^{\nu}
\quad &=\quad\varepsilon_{abc}F^{c}{}_{\mu\nu}
\delta A^{a\nu}\wedge\delta A^{b\mu}\cr
\quad &=\quad 0. & (47)\cr
\noalign{\hbox{Next we consider}}
tr\enspace\delta F_{\mu\nu}\wedge\nabla^{\mu}\delta A^{\nu}
\quad &=\quad{1\over 2} tr\enspace\delta F_{\mu\nu}\wedge
\left(\nabla^{\mu}\delta A^{\nu} -\nabla^{\nu}\delta 
A^{\mu}\right)\cr
\quad &=\quad{1\over 2} tr\enspace\delta F_{\mu\nu}\wedge
\delta F^{\mu\nu}\quad =\quad 0. & (48)\cr
}$$
\noindent Combining (47) and (48), we see that
$$\openup 2mm
\partial^{\mu}J_{\mu}\quad =\quad 0\eqno (49)
$$
\noindent by (44). It follows that the closed functional 
$2$-form, $\Omega$, given by
$$\openup 2mm
\Omega\quad =\quad\int_{\Sigma_t}d\sigma^{\mu} J_{\mu}
\quad =\quad\int_{\Sigma_t}d\sigma^{\mu}\enspace
tr\enspace\delta F_{\mu\nu}\wedge\delta A^{\nu},\eqno (50)
$$
\noindent is Lorentz invariant. Thus we have constructed a 
Lorentz invariant and gauge invariant symplectic form, $\Omega$, 
on the $SU(2)$ Yang-Mills phase space.
 
We can obtain the standard $SU(2)$ Yang-Mills symplectic form by 
a suitable choice of the space-like hyper-surface, $\Sigma_{t}$ :
$$\openup 2mm
\Omega\quad =\quad\int_{\Sigma_t}{d^3}x\enspace tr
\enspace\delta E_{i}\wedge\delta A^{i},\eqno (51)
$$
\noindent where $E_{i} = F_{0i}$ is the momentum canonically 
conjugate to ${A^i}$.

\vskip 5mm
\noindent {\bf 4. A Symplectic Form For Ashtekar's Canonical 
Gravity}
\vskip 5mm
 
The inverse densitized triads and the Ashtekar connection act 
as symplectic co-ordinates in the phase space of Ashtekar's 
canonical gravity. We wish to put a symplectic form on Ashtekar's 
phase space in a manner consistent with the Crnkovic-Witten 
construction. An extra difficulty here, over and above the 
problem of gauge invariance, is the complex nature of Ashtekar's 
canonical variables. Denoting the functional exterior derivative 
of forms on Ashtekar's phase space by $\delta$, we have
$$
\delta A_{ai}\quad =\quad{\delta\omega_{0ai}\over\delta E^{bj}}
\delta E^{bj} + {\delta\omega_{0ai}\over\delta{\dot E}^{bj}}
\delta{\dot E}^{bj} - {i\over 2}\varepsilon_{acd}
{\delta\omega^{cd}{}_{i}\over\delta E^{bj}}\delta E^{bj},
\eqno (52)
$$
where the shorthand notation ${\delta\omega_{0ai}\over
\delta E^{bj}}
=\int_M{\delta\omega_{0ai}(x)\over\delta E^{bj}(y)}d^4y $ is 
understood.
 
We require the symplectic form,
$$\openup 2mm\eqalignno{
\Omega\quad &=\quad\int_{\Sigma_t}{d^3}x\enspace
{\delta}E^{ai}\wedge{\delta}A_{ai}\cr
\quad &=\quad -\int_{\Sigma_t}{d^3}x\left(
{\delta A_{ai}\over\delta E^{bj}}\delta E^{bj}\wedge\delta E^{ai}
+{\delta A_{ai}\over\delta{\dot E}^{bj}}\delta{\dot E}^{bj}
\wedge\delta E^{ai}\right), & (53)\cr
}$$
\noindent to be real-valued in order to have a unique symplectic 
structure on Ashtekar's phase space. A complex-valued symplectic 
form would give rise to two real symplectic structures. Moreover, 
a real-valued symplectic form produces real-valued Poisson 
brackets.
 
Working in the time gauge and using (6) and (7), it is 
straightforward to show that
$$\openup 2mm\eqalignno{
&\int_{\Sigma_t}{d^3}x{\delta\omega_{0ai}\over\delta E^{bj}}
\delta E^{bj}\wedge\delta E^{ai}
\quad =\quad{1\over 2{\cal N}}
\Bigl [\bigl ((E^{-1})_{bi}(E^{-1})_{ak} 
+ (E^{-1})^{c}{}_{i}(E^{-1})_{ck}\delta_{ab}
\bigr )\partial_{j} N^{k}\cr
&-\bigl ((E^{-1})_{bi}(E^{-1})^{c}{}_{j}(E^{-1})_{ck} 
+ (E^{-1})_{bk}(E^{-1})^{c}{}_{i}(E^{-1})_{cj}\bigr )
\bigl (\dot{E}_{a}{}^{k} + E_{a}{}^{\ell}\partial_{\ell}N^{k}
- N^{\ell}\partial_{\ell} E_{a}{}^{k}\bigr)\cr
&- (E^{-1})_{ai}(E^{-1})_{cj}(E^{-1})_{bk}\bigl (
N^{\ell}\partial_{\ell} E^{ck} -\dot{E}^{ck}\bigr )\Bigr ]
\delta E^{bj}\wedge\delta E^{ai}, & (54)\cr
&\int_{\Sigma_t}{d^3}x{\delta\omega_{0ai}\over\delta\dot{E}^{bj}}
\delta\dot{E}^{bj}\wedge\delta E^{ai}\quad =\quad\cr
&{1\over 2{\cal N}}
\bigl [(E^{-1})^{c}{}_{i}(E^{-1})_{cj}\eta_{ab} + (E^{-1})_{bi}
(E^{-1})_{aj}
- (E^{-1})_{ai}(E^{-1})_{bj}\bigr ]
\delta\dot{E}^{bj}\wedge\delta E^{ai}, & (55)\cr
}$$
 
Following Henneaux {\it et al} [7], we can write 
$$\openup 2mm\eqalignno{
\varepsilon_{acd}\omega^{cd}{}_{i}\quad &=\quad
{\delta\over\delta E^{ai}}\int_{\Sigma_t}{d^3}x\enspace G,\cr
\noalign{\hbox{where}}
G\quad &=\quad\varepsilon^{jk\ell} h_{bj}
\partial_{k} h^{b}{}_{\ell}. & (56)\cr
}$$
\noindent Then
$$\openup 2mm\eqalignno{
\int_{\Sigma_t}{d^3}x\enspace\varepsilon_{acd}
{\delta\omega^{cd}{}_{i}\over\delta E^{bj}}
\delta E^{bj}\wedge\delta E^{ai}\quad &=\quad
{\delta\over\delta E^{bj}}\Biggl({\delta\over\delta E^{ai}}
\int_{\Sigma_t}{d^3}x\enspace G\Biggr)
\delta E^{bj}\wedge\delta E^{ai} & (57)\cr
\quad &=\quad -{\delta\over\delta E^{bj}}\Biggl
({\delta\over\delta E^{ai}}\int_{\Sigma_t}{d^3}x\enspace G\Biggr)
\delta E^{bj}\wedge\delta E^{ai}\quad =\quad 0.\cr
}$$
\noindent Thus the complex part of the symplectic form, $\Omega$, 
is zero in the time gauge.
 
Now we must show that the symplectic form is real-valued for 
all other gauges, apart from the time gauge. When we go from a 
time 
gauge hyper-surface to a more general hyper-surface, the  
group of local symmetries enlarges from the rotation group, 
$SO(3)$, to the 
Lorentz group, $SO(3, 1)$ (see the appendix).

Ashtekar's action can be written :
$$\openup 2mm\eqalignno{
S\quad &=\quad\int_M{\cal L}{d^4}x,&\cr
\noalign{\hbox{where}}
{\cal L}{d^4}x\quad &=\quad iF_{a}\wedge\Lambda^{a}, & (58)\cr
}$$
\noindent as in (17) and (20). The symplectic form (53) can be then be written :
$$\openup 2mm
\Omega\quad =\quad\int_{\Sigma_t}{d^3}x\enspace\delta\left(
{\delta{\cal L}\over\delta{\dot A}_{ai}}\right)
\wedge{\delta}A^{ai}.\eqno (59)
$$

The analogy with the Hamiltonian formulation of the $SU(2)$ 
Yang-Mills field suggests that we ought to postulate a functional 
$2$-form on Ashtekar's phase space, with the vector density on 
space-time :
$$\openup 2mm\eqalignno{
J_{\mu}\quad =\quad 
\delta\left [{\delta{\cal L}
\over\delta({\partial_\mu}A_{ai})}\right ]
\wedge\delta A^{ai}
\quad &=\quad\delta\left [{\delta{\cal L}\over
\delta({\partial_\mu}A_{a\nu})}\right ]\wedge\delta A^{a\nu}\cr
\quad &=\quad i\enspace
tr\enspace\delta\Lambda_{\mu\nu}\wedge\delta A^{\nu}, & (60)\cr
}$$
\noindent as a symplectic current. Here, the trace $tr$ relates 
to a representation of $sl (2, {\bf C})$, the Lie algebra of 
$SO(3, 1)$. We can associate an $SO(3, 1)$ co-variant derivative, 
$D_{\mu}$, with the connection, $A_{\mu}$, such that
$$\openup 2mm\eqalignno{
\delta F_{\mu\nu}\quad &=\quad D_{\mu}\delta A_{\nu}
- D_{\nu}\delta A_{\mu}. & (61)\cr}
$$
 
Under a local $SO(3, 1)$ transformation, 
$A_{\mu}\to A_{\mu} +\partial_{\mu}\lambda + [A_{\mu},\lambda ]$, 
where $\lambda$ is a real-valued function on space-time. It is 
found that 
$\delta A_{\mu}\to\delta A_{\mu} + [\delta A_{\mu},\lambda ]$ and 
$\delta\Lambda_{\mu\nu}\to\delta\Lambda_{\mu\nu} + 
[\delta  \Lambda_{\mu\nu},\lambda ]$. It follows that the 
symplectic current is an $SO(3, 1)$-singlet. Thus we can write :
$$\openup 2mm\eqalignno{
\partial^{\mu} J_{\mu}\quad &=\quad 
D^{\mu} J_{\mu}& (62)\cr
\quad &=\quad i\enspace tr\enspace D^{\mu}\delta\Lambda_{\mu\nu}
\wedge\delta A^{\nu} + i\enspace tr\enspace\delta\Lambda_{\mu\nu}
\wedge D^{\mu}\delta A^{\nu}. & (63)\cr
}$$
\noindent Since
$$\openup 2mm\eqalignno{
D^{\mu}\Lambda_{\mu\nu}\quad &=\quad 0, & (64)\cr
\noalign{\hbox{we have}}
D^{\mu}\delta\Lambda_{\mu\nu}\quad &=\quad   
- [\delta A^{\mu}, \Lambda_{\mu\nu}]&\cr
\noalign{\hbox{and}}
i\enspace tr\enspace D^{\mu}\delta \Lambda_{\mu\nu}
\wedge\delta A^{\nu}\quad &=\quad 
-i\enspace tr\enspace [\delta A^{\mu},\Lambda_{\mu\nu}]
\wedge\delta A^{\nu}\cr
\quad &=\quad{i\over 2}tr\enspace [\delta A^{\mu},\delta A^{\nu}]
\wedge\Lambda_{\mu\nu}\quad =\quad 0. & (65)\cr
}$$
 
When we vary the action with respect to the orthonormal  
$1$-forms, while keeping the connection $1$-forms fixed, 
the equations of motion imply:
$$\openup 2mm
F^{\mu\nu}\delta\Lambda_{\mu\nu}\quad =\quad 0.\eqno (66)
$$
\noindent Consequently,
$$\openup 2mm
i\enspace tr\enspace\delta\Lambda_{\mu\nu}\wedge D^{\mu}
\delta A^{\nu}
= {i\over 2}\enspace tr\enspace\delta\Lambda_{\mu\nu}
\wedge\delta F^{\mu\nu} =
-{i\over 2}\enspace tr\enspace\delta
\left(F^{\mu\nu}\delta\Lambda_{\mu\nu}\right)= 0.\eqno (67)
$$

It is clear from (65) and (67) that the divergence of the 
symplectic current in (63) vanishes, and it follows that the 
closed functional $2$-form 
$$\openup 2mm
\Omega\quad =\quad\int_{\Sigma_{t}}d{\sigma^\mu} J_{\mu}\quad =
\quad i\int_{\Sigma_{t}}d{\sigma^\mu}\enspace tr\enspace\delta 
\Lambda_{\mu\nu}\wedge\delta A^{\nu},\eqno (68)
$$
\noindent is a Lorentz invariant and general co-ordinate 
invariant symplectic form on Ashtekar's phase space. In 
particular, since the imaginary part of $\Omega$ vanishes 
in the time gauge, and $\Omega$ is Lorentz invariant, then 
$\Omega$ is {\it real-valued} in {\it any} local Lorentz frame, 
even one in which the inverse densitized triads are complex-valued.
\vfill\eject
\noindent {\bf 5. Conclusions}
\vskip 5mm

We have described Ashtekar's canonical gravity in a manifestly 
co-variant way by using a construction due to Crnkovic and Witten.
This construction had worked for the ADM formulation of general 
relativity, so we hoped it might work for Ashtekar's formulation, 
at least in the time gauge. The only obstacles to be overcome 
were gauge invariance, and the complex nature 
of Ashtekar's canonical variables.

Gauge invariance was incorporated into the symplectic form we 
put on Ashtekar's phase space, along with Lorentz invariance, 
as in   
the work of Crnkovic and Witten. Using a result in a paper by 
Henneaux {\it et al}, we showed that the symplectic form is 
{\it real-valued} in the time gauge, thereby giving rise to a 
{\it unique} symplectic structure on Ashtekar's phase space, as 
well as real-valued Poisson brackets.

It remained to show that the symplectic form is real-valued for 
all other gauges, in addition to the time gauge, when the 
canonical variables are all $sl (2, {\bf C})$-valued. This was 
accomplished using the analogy with the Hamiltonian formulation 
of the $SU(2)$ Yang-Mills field. As a result, we know that the 
Crnkovic-Witten construction can be applied to Ashtekar's 
canonical gravity.
\vskip 5mm
\noindent {\bf Appendix}
\vskip 5mm

A general orthonormal basis can be obtained from one adapted to 
a space-like hyper-surface, $\Sigma_{t}$, as follows. Denoting 
4-dimensional orthonormal and co-ordinate indices by $A, B,
\dots$ and $\mu,\nu,\dots$ respectively, let
$$\openup 2mm
h^{^{A}}{}_{\mu}\quad =\quad\left [\matrix{N & 0\cr
h^{a}{}_{j} N^{j} & h^{a}{}_{i}\cr}\right]\eqno (69)
$$
\noindent be a tetrad with $h^{0} = N d{x^0}$ normal to 
$\Sigma_{t}$. This choice of tetrad is compatible with the 
time gauge condition. Here $N$ is the lapse function, $N^{i}$ 
are the shift functions, and $h^{a}{}_{i}$ is an orthonormal 
triad on $\Sigma_{t}$ satisfying
$$\openup 2mm
h^{a}{}_{i}h_{aj}\quad =\quad g_{ij},\qquad
(h^{-1})_{a}{}^{i}h^{a}{}_{j}\quad =\quad\delta^{i}{}_{j},
\eqno (70)
$$
\noindent where $g_{ij}$ is the 3-dimensional metric on 
$\Sigma_{t}$. An arbitrary Lorentz boost, tangent to 
$\Sigma_{t}$, with 
3-velocity, $v^{a}$, is given by
$$\openup 2mm\eqalignno{
L(v)^{^{A}}{}_{_{B}}\quad &=\quad\left [\matrix{\gamma 
& -\gamma v_{b}\cr
-\gamma v^{a} & \delta^{a}{}_{b} 
+ {{\gamma^2}\over{1 +\gamma}}{v^a}{v_b}\cr}\right],\cr
\noalign{\hbox{where}}
\gamma\quad &=\quad\left (1 - {v^a}{v_a}\right)^{-{1\over 2}}. 
& (71)\cr
\noalign{\hbox{So an arbitrary tetrad is of the form :}} 
e^{^{A}}{}_{\mu}\quad &=\quad L(v)^{^{A}}{}_{_{B}} 
h^{^{B}}{}_{\mu}. & (72)\cr
\noalign{\hbox{Note, however, that}}
e^{a}{}_{i}\quad &=\quad h^{a}{}_{i} + 
{{\gamma^2}\over{1 +\gamma}}{v^a}{v_b}h^{b}{}_{i} & (73)\cr
}$$
\noindent are not an orthonormal {\it triad} because 
$e^{a}{}_{i}e_{aj}\ne g_{ij}$. Let $e$ be the determinant 
of the matrix, $[e_{ai}]$. Defining the inverse densitized 
triads, $E^{ai}$, by
$$
E^{ai}\quad =\quad e (e^{-1})^{ai}
- i\varepsilon^{ijk}e^{a}{}_{j}e^{b}{}_{k}v_{b},\eqno (74)
$$
 
Ashtekar's Lagrangian takes the form :
$$\openup 2mm\eqalignno{
L\quad =\quad\int_{\Sigma_{t}}{d^3}x
\biggl [&A_{ai}\dot{E}^{ai}
- A_{a0}\bigl (\partial_{i} E^{ai}
- i\varepsilon^{abc}A_{bi}E_{c}{}^{i}\bigr )\cr
& +{i\over 2}{\cal N}\varepsilon^{abc}
F_{aij}E_{b}{}^{i}E_{c}{}^{j} 
+ N^{i}F_{aij}E^{aj}\biggr ], & (75)\cr
}$$
\noindent where ${\cal N} = \gamma N/e$. This shows that $E^{ai}$ 
and $A_{ai}$ are canonically conjugate, with ${\cal N}, N^{i}$, 
and 
$A_{a0}$ behaving as Lagrange multipliers for the secondary 
constraints. It is straightforward to verify that the complex 
inverse densitized triads satisfy
$$
E^{ai}E_{a}{}^{j}\quad =\quad gg^{ij},\qquad
E^{ai}E_{b}{}^{j}g_{ij}\quad =\quad g\delta^{a}{}_{b},
\eqno (76)
$$
\noindent where $g$ is the determinant of the matrix, 
$[g_{ij}]$, and so they can be regarded, in a sense, as a 
complex orthonormal triad density. The effect of an infinitesimal 
Lorentz boost on the canonical variables is easily calculated. 
Using
$$\openup 2mm\eqalignno{
\delta L^{^{A}}{}_{_{B}}(0)\quad &=\quad\left [\matrix{0 
& -\delta{v_b}\cr
-\delta{v^a} & 0}\right], & (77)\cr
\noalign{\hbox{we find}}
\delta_{v}A^{a}{}_{i}\quad &=-(\partial_{i}\delta v^{a}
- i\varepsilon^{abc}A_{bi}\delta v_{c}), & (78)\cr
\delta_{v}E^{ai}\quad &=\quad i\varepsilon^{abc}E_{b}{}^{i}
\delta v_{c}, & (79)\cr
}$$
\noindent which is to be compared with the effect of an 
infinitesimal tangent space rotation on $\Sigma_{t}$, 
parameterised by $\delta\theta^{a}$,
$$\openup 2mm\eqalignno{
\delta_{\theta}A^{a}{}_{i}\quad &=\quad i
\left (\partial_{i}\delta \theta^{a} - i\varepsilon^{abc}A_{b}
{}_{i}\delta\theta_{c}\right), & (80)\cr
\delta_{\theta}E^{ai}\quad &=\quad \varepsilon^{abc}E_{b}{}^{i}
\delta\theta_{c}. & (81)\cr
}$$

As an extra check that these variables are canonically 
conjugate, it is instructive to prove that Lorentz 
transformations leave the Poisson brackets unchanged. It is 
easy to verify that an infinitesimal boost leaves the Poisson 
bracket unchanged as, of course, do infinitesimal rotations. As 
boosts and rotations form a group, we can simply exponentiate 
and deduce that {\it finite} Lorentz transformations also leave 
the Poisson bracket invariant.
Hence,
$$
\left\{E^{ai}, A_{bj}\right\}\quad =\quad 
\delta^{a}{}_{b}\delta^{i}{}_{j}\eqno (82)
$$
\noindent must hold for the complex $E^{ai}$ with $v^{a}\ne 0$. 
In conclusion, it has been shown that it is not necessary to 
match the choice of an orthonormal frame to the foliation of 
space-time in Ashtekar's canonical gravity. The inverse 
densitized triads are now complex-valued, but there are still 
conditions on them, since the imaginary part only has three 
degrees of freedom, ${v^a}$, rather than the nine which would 
be necessary for a complex $3\times 3$ matrix. The Ashtekar 
connection, $A_{ai}$, become $sl (2, {\bf C})$-valued. The 
infinitesimal $sl (2, {\bf C})$ gauge transformations are 
given above in (78) and (79). Finally, this appendix is 
equivalent to the work of Ashtekar {\it et al} in [10], 
where the results are formulated in spinor notation.
\vfill\eject

\noindent {\bf References}
\vskip 1cm
\openup 2mm
\+ & [1]. A. Ashtekar, Phys. Rev. Lett. {\bf 57}, 2244 (1986);
Phys. Rev. D {\bf 36}, 1587 (1987).\cr
\+ & [2]. J.N. Goldberg, Phys. Rev. D {\bf 37}, 2116 (1988).\cr
\+ & [3]. C.J. Isham, in {\it Superstrings and Supergravity}, 
ed. A.T.Davies and D.G.Sutherland,\cr
\+ & Proceedings of the $\hbox{$28$}^{th}$ Scottish Universities 
Summer School in Physics (1985).\cr
\+ & [4]. E.Witten, Nucl.Phys. B {\bf 276}, 291 (1986);\cr
\+ & G.Zuckerman, 
in {\it Mathematical Aspects of String Theory},\cr
\+ & ed. S.T.Yau, World Scientific (1987).\cr
\+ & [5]. C.Crnkovic and E.Witten, in {\it Three Hundred Years 
of Gravitation},\cr
\+ & ed. S.W.Hawking and W.Israel, Cambridge University 
Press (1987).\cr 
\+ & [6]. J.N. Goldberg and C. Soteriou, 
Class.Quant.Grav. {\bf 12}, 2779 (1996).\cr
\+ & [7]. M.Henneaux, J.E.Nelson, and C.Schomblond, 
Phys.Rev. D {\bf 39}, 434 (1989).\cr
\+ & [8]. T.Jacobson and L.Smolin, 
Class.Quant.Grav.{\bf 5}, 583 (1988)\cr
\+ & [9]. A. Ashtekar, in {\it New Perspectives in 
Canonical Gravity}, Bibliopolis (1988). \cr
\+ & [10]. A.Ashtekar, A.P.Balachandran, S.Jo, Int.J.Mod.Phys. A 
{\bf 4}, 1493 (1989).\cr
\vfill\eject

\end